## Learning Rules for Materials Properties and Functions

Mario Boley[1] and Matthias Scheffler[2], 1) Department of Data Science and AI, Monash University, Melbourne, Australia, 2) The NOMAD Laboratory at the FHI-MPG and HU, Berlin, Germany

**Status**

In materials science and engineering, one is typically searching for materials that exhibit exceptional performance for a certain function, and the number of these materials is extremely small. Thus, statistically speaking, we are interested in the identification of "rare phenomena", and the scientific discovery typically resembles the proverbial hunt for the needle in a haystack. Let us illustrate this with a "classical" example, i.e. searching for materials that are very robust, highly transparent, and at the same time have a high heat conductivity. In the immense space of structural and chemical materials, there is one strong high-performance candidate: carbon in the diamond structure. Hardly any other material comes close. And from a thermodynamic perspective, this material is not even stable but metastable. As we understand the mechanisms behind the mentioned properties, we trust the conclusion that diamond is the exceptional champion of the issued search. But how can we reliably find materials that exhibit exceptional performance for functions in general, for example, for catalysis, photovoltaics, or batteries? All searches face the following situation[1]:

- The number of possible materials is practically infinite.
- The electronic and atomistic processes that rule a desired materials function are many, and their concerted action is typically highly complex and intricate, resulting an immense number of possibly relevant mechanisms.
- The number of data that are "clean" (comprehensively characterized and high-quality) and *relevant* for the function of interest are typically very low.

Under these daunting conditions we aim to identify the *rules* that govern the rare phenomena corresponding to particularly exceptional materials. Such rules describe regions in materials spaces that are relevant for the function of interest (see Fig. 1). In analogy to biology, the basic physico-chemical parameters entering these rules may be called "materials genes", as they are related to processes that trigger, actuate, or facilitate, or hinder the property of interest. In particular, we are interested in such regions that 1) contain exclusively or at least predominantly materials with desired properties and 2) are described in a way that allows us to efficiently sample from them new synthesizable materials. Publicly shared materials databases and AI methods have enabled encouraging progress[1] towards this goal (see Fig. 2 as an example)[2]. However, critical challenges remain.

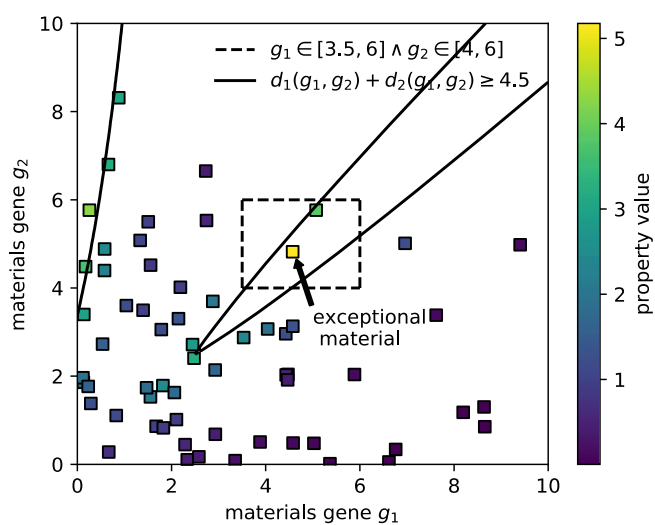

**Figure 1.** By mapping materials (depicted as squares) into spaces defined by relevant "materials genes" we can identify regions where materials exhibit desired exceptional properties. Depending on the employed AI methods, such regions can be given by simple Boolean conditions (dashed line) or in terms of more complex analytic functions of the genes (solid line). Typically, the space of relevant genes will have a much higher dimensionality than two.

**Current and Future Challenges**

Most available data science and machine learning methods are fundamentally unsuited for the required identification of rare phenomena. Firstly, they typically aim to fit a global model to the available data by minimizing the "regularized" average error. This focus on average global performance not only puts importance on accurately modelling the hay instead of the needles. Even worse, regularization means to deliberately avoid modelling the extra-ordinary for the sake of avoiding overfitting. Secondly, as pointed out by Ghiringhelli et al.[3], off-the-shelf methods cannot reliably identify meaningful and trustworthy rules that describe exceptional materials, because they implicitly or explicitly rely on descriptors (also called representations) of materials that are either too restrictive (because they are hand-picked) or too unrestrictive (e.g., in the case of deep learning) and thus model "non-physical" relations likely unrelated to the materials genes of relevance.

Using symbolic regression and compressed sensing, the SISSO (sure independence screening and sparsifying operator) approach[4] alleviates this problem by identifying descriptors consisting of typically only a few analytical functions of relevant materials genes. Based on its physical plausibility and robust empirical performance, we can say with some confidence that this approach successfully identifies rules satisfying our first criterion: the description of regions that predominantly contain desired materials. A remaining problem lies in the second requirement: our ability to efficiently sample interesting novel materials. Rejection sampling can be employed to generate candidates if the considered materials class is small, e.g., binary systems restricted to a few crystal structures. However, this does not scale to the vast design spaces relevant for general searches. The central challenge is that SISSO similar to other commonly used descriptors are not efficiently invertible. While representing materials through their genes enables us to discover reliable rules, many points in gene space do not correspond to real materials, and this complicates the direct generation of new candidates from a specific region.

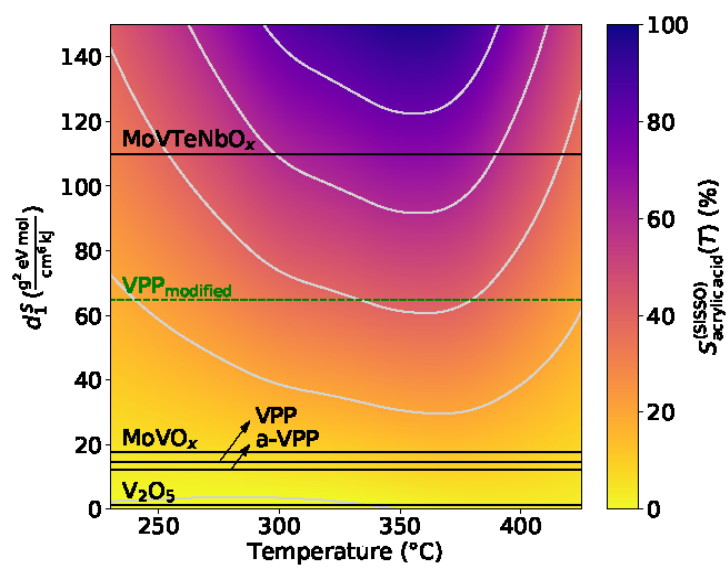

Figure 2. Map of catalysts given by the SISSO model for the selectivity of propane conversion to acrylic acid. For details see Ref. 2. The desired high selectivity situation is colored dark blue. The materials used for deriving the descriptors are indicated by the black lines. The function $d_1^S$ is quite complex, identifying the macroscopy material, i.e. its basic properties (e.g. composition) as well as its porosity etc. VPP$_{modified}$ is a suggested material that would result by changing some materials parameters of the VPP material. In general, however, the relationship "real materials" $\rightarrow d_1^S$ is not invertible.

**Advances in Science and Technology to Meet Challenges**

An important alternative approach to rule identification is subgroup discovery (SGD)[5]. Similar to SISSO, SGD also describes non-linear relations between materials genes and properties. However, in contrast to SISSO, the SGD rules are given as Boolean conjunctions of conditions on individual genes. This means that the described regions in gene space are simple axis-parallel (hyper-)rectangles, which makes it easier to generate novel materials from them: one simply has to satisfy the given conditions on each relevant gene independently. However, as above, many, probably most, combinations of the gene values may not correspond to real materials.

Furthermore, currently available SGD methods are, unfortunately, not designed to describe rare phenomena. They are based on ideas from confirmatory statistics (significance testing) to derive final conclusive results from a given dataset. To assure results that are significant for the data at hand, they prioritize the detection of relatively frequent phenomena. Fortunately, in the context of materials science, this extremely conservative approach of one-shot correctness can be relaxed. Since we have computational methods that can obtain accurate new data with reasonable efficiency, we can aim for an approach where pattern discovery and first-principles methods work in unison to facilitate rapid scientific discovery.

Borrowing ideas from Thompson sampling and Bayesian optimization[6], such rule discovery methods should propose rules that are reasonable candidates to describe the rare material champions and then obtain new simulated data from the proposed regions to validate or falsify this proposal. By repeating this process, we iteratively arrive at new regions where desired materials are more and more likely to be found. Instead of one-shot correctness, this approach aims to identify the desired rare phenomenon as soon as possible in this iterative process by optimizing an exploration/exploitation trade-off[a].

This compelling vision provides a clear agenda of statistical and algorithmic problems to tackle: Firstly, we need a sound selection mechanism for hypotheses about rare phenomena that appropriately compromises between the value of a rule and the likelihood that it can be confirmed by future data. Secondly, we need efficient algorithms that find optimal regions based on this selection mechanism.

**Concluding Remarks**

In summary, publicly shared materials data and AI code, as provided by the NOMAD AI Toolkit[7], as well as physically plausible representations based on materials genes (like the ones used in SISSO and SGD) have facilitated progress towards identifying rules that describe desired materials. So far, however, all approaches are lacking either the ability to consistently describe only promising materials or the ability to efficiently generate them – at least at the ultimately required scale. To advance further, challenging statistical and algorithmic problems have to be solved, but there are promising starting points: The combination of Bayesian approaches to multiple hypothesis testing[8] as well as the versatile branch-and-bound approach[9] to discrete optimization stands a good chance to enable the envisioned methods. However, due to their reliance on adaptively generated new data, their development will require a concentrated interdisciplinary effort between materials and data science.

---

[a] Here, "exploration" refers to sampling from regions where one is still uncertain about materials performance, and "exploitation" refers to sampling from regions with relatively strong and certain materials performance.

**Acknowledgements**

*We acknowledge Luca Ghiringhelli, Lucas Foppa, Claudia Draxl, Wray Buntine, and Daniel Schmidt for insightful discussions and, in particular, thank Luca Ghiringhelli and Lucas Foppa for critically reading the manuscript. This work received funding from the European Union's Horizon 2020 Research and Innovation Programe (grant agreement Nº 951786), the NOMAD CoE, and ERC: TEC1P (No. 740233) as well as from the Australian Research Council (DP210100045).*

**References**

[1] Draxl C and Scheffler M 2020 Big-data-driven materials science and its FAIR data infrastructure Handbook of Materials Modeling ed S Yip et al (Basel: Springer International Publishing) p. 49 DOI: 10.1007/978-3-319-44677-6_104

[2] Foppa L Ghiringhelli L M Girgsdies F Hashagen M Kube P Hävecker M Carey S Tarasov A Kraus P Rosowski F Schlögl R Trunschke A and Scheffler M 2021 Materials genes of heterogeneous catalysis from clean experiments and artificial intelligence submitted (preprint https://arxiv.org/abs/2102.08269)

[3] Ghiringhelli L M Vybiral J Levchenko S V Draxl C and Scheffler M 2015 Big data of materials science: critical role of the descriptor Phys. Rev. Lett. **114**, 105503 DOI: 10.1103/PhysRevLett.114.105503

[4] Ouyang R Curtarolo S Ahmetcik E Scheffler M and Ghiringhelli L M 2018 SISSO: a compressed-sensing method for identifying the best low-dimensional descriptor in an immensity of offered candidates Phys. Rev. Mat. **2**, 083802

[5] Goldsmith B R Boley M Vreeken J Scheffler M Ghiringhelli LM 2017 Uncovering structure-property relationships of materials by subgroup discovery New J. Phys. **19**, 013031 DOI: 10.1088/1367-2630/aa57c2.

[6] Frazier P I and Wang J 2016 Bayesian optimization for materials design Information Science for Materials Discovery and Design ed Lookman T et al (Springer, Cham.) pp 45-75 DOI 10.1007/978-3-319-23871-5_3

[7] NOMAD Artificial Intelligence Toolkit https://nomad-lab.eu/index.php?page=AItutorials

[8] Scott JG Berger JO 2006 An exploration of aspects of Bayesian multiple testing Journal of statistical planning and inference 136(7) pp 2144-2162 DOI 10.1016/j.jspi.2005.08.031

[9] Boley M Goldsmith BR Ghiringhelli LM Vreeken J 2017 Identifying consistent statements about numerical data with dispersion-corrected subgroup discovery. Data Mining and Knowledge Discovery, 31(5) pp 1391-1418 DOI 10.1007/s10618-017-0520-3.